**Kevin Geyer Harrison, Department of Earth and Environmental Sciences, Denison University, Granville, OH 43023 USA**
KevinHarrison44@gmail.com
**Orcid ID # 0000-0002-6921-7203**
www.KevinGeyerHarrisonPhD.org

**ABSTRACT**

Removing carbon dioxide from the atmosphere may slow climate change and ocean acidification. My approach converts atmospheric carbon dioxide into graphite (CD2G). The net profit for this conversion is ~$381/ton $CO_2$ removed from the atmosphere. At the gigaton scale, CD2G factories will increase the affordability and availability of graphite. Since graphite can be used to make thermal batteries and electrodes for fuel cells and batteries, CD2G factories will help lower the cost of storing renewable energy, which will accelerate the transition to renewable energy. Replacing fossil fuel energy with renewable energy will slow the release of carbon dioxide to the atmosphere, also slowing climate change. Converting atmospheric carbon dioxide into graphite will both generate a profit and slow climate change.

**INTRODUCTION**

Increasing levels of atmospheric carbon dioxide may cause climate change and prove costly. Multiple researchers have estimated the costs associated with increasing atmospheric carbon dioxide levels. Rennert et al. (2022) have estimated that the social costs for a ton of carbon dioxide released to the atmosphere range from $44 to $413. Kikstra et al. (2021) have estimated that the social cost of carbon dioxide is $307/ton. Archer et al. (2020) have estimated that the "ultimate" cost of carbon dioxide is ~$100,000/ton. Some of the social costs of additional carbon dioxide include their impact on infectious disease (Kupferschmidt, 2023). It would be hard to overestimate the cost of the damages associated with increasing atmospheric carbon dioxide levels. For example, Wong (2023) has reported that the 12 months preceding November 2023 were the hottest on record and were 1.32 °C. above pre-industrial temperatures. Rodrigues (2023) has linked extreme drought in the Amazon rainforest with climate change. Carbon dioxide levels reached record highs of about 421 ppm in May of 2022 (Fountain, 2022). Coal-fired power plants release particulate matter that kills people (Henneman et al., 2023, Mendelsohn & Kim, 2023), along with releasing a significant amount of carbon dioxide to the atmosphere (Houghton, 2004). Replacing coal-fired plants with renewable energy will save lives, in addition to slowing climate change.

Two ways to decrease the chances of catastrophic climate change are by switching to renewable energy and using Carbon Capture, Utilization, and Storage (CCUS). My approach uses CCUS to both remove carbon dioxide from the atmosphere and to accelerate the transition to renewable energy by converting carbon dioxide to graphite (CD2G) using Direct Air Capture (DAC; Fig. 1). The CD2G approach removes carbon dioxide from the atmosphere via adsorption and converts it into graphite using the Bosch reaction (Fig. 1). Lin et al. (2021) have developed a metal-organic framework to capture carbon dioxide, and Zhou et al. (2021) made an iron-containing mordenite monolith to trap carbon dioxide. When the carbon dioxide is released, it is converted into graphite (Fig. 1).

Graphite is essential for the renewable energy transition, which requires energy storage. For example, carbon blocks can be used to store renewable energy as thermal energy (Ramkumar & Patterson, 2024). Graphite bocks may serve the same purpose. Graphite is already used for fuel cell electrodes (Rao et al., 2020) and lithium-ion battery electrodes (Gastol, et al., 2021). Since thermal batteries, fuel cells and lithium-ion batteries can store renewable energy, increasing the availability and affordability of graphite will accelerate the transition to renewables by decreasing the cost and increasing the availability of energy storage. The amount of graphite needed for batteries is large. For example, electric vehicles need 50 to 100 kgs of graphite on average (Lienert & Carey, 2023).

Currently, graphite is in short supply and the demand for graphite is growing faster than the supply (Ballinger et al., 2019). Prices for flake graphite range from $500/ton to $2,300/ton (Table 1). The increased demand for graphite coupled with reduced availability may increase the price of graphite and slow the transition to renewable energy. Graphite is used by metal, chemical, electrical, nuclear, and rocket industries (Moore & Volk, 2020). For example, graphite electrodes are used in electric furnaces to make steel; graphite anodes are used in the electrolytic production of various substances, such as hydrogen; graphite is used to make motors/generator brushes, seals, bearings, nozzles for rocket motors, metallurgical molds and crucibles, reaction vessel linings, heat exchangers, pumps, pipings, and valves (Moore & Volk, 2020). Graphite electrodes can be used for generating hydrogen (Yuvaraja & Santhanaraj, 2014).

*Graphene*

Graphite can be used to make graphene (Hernandez et al., 2008; Achee et al., 2018). Making graphite more affordable will also increase the affordability and availability of graphene, which has great potential to play a key role in advancing technology and solving societal problems (Allen et al., 2010; Tiwari et al., 2020). Graphene prices range from ~$100 to ~$400/gram (~$100,000,000 to ~$400,000,000/metric ton) (https://investingnews.com/daily/tech-investing/nanoscience-investing/graphene-investing/graphene-cost/; accessed 8/6/24). Graphene may function as a zero-bandgap semiconductor, have applications in ultrafast photonics and supercomputers, thermal management, batteries, displays, structural composites, and catalyst supports (Tiwari et al., 2020). Graphene can be used to make metal organic frameworks (Zhang et al., 2022), catalysts (Zhao et al., 2022), molecular sieves (Huang et al., 2021), for desalination (Boretti et al., 2018; Liang et al., 2021), to split water to produce hydrogen and oxygen efficiently (Raj et al., 2022; Wang et al., 2021), as a supercapacitor (Tiwari et al., 2020, Zaka et al., 2021, Wang et al., 2022), in nanotechnology (Wang et al., 2023), as a superconductor (Zhou et al., 2022; Kim et al., 2023), for solid state batteries (Pervez et al., 2022), to improve battery performance (Khan et al., 2023), as a microchip/semiconductor (Zhao et al., 2024; Iocopi & Ferrari, 2024), for quantum semiconductors (Assouline et al., 2023),for fuel cells (Zhao et al., 2022), and carbon nanotubes made from graphene can scavenge waste heat (Dyatkin, 2021). Graphene-copper materials have a 450% higher electrical current carrying capacity, 41% higher electrical conductivity, and 224% higher thermal heat dissipation compared to pure copper (https://mae.osu.edu/events/2023/03/can-graphene-based-electrical-conductors-replace-copper; 8/24/24). This increased performance could improve the ability of the grid to move renewable energy long distances. Thus, increasing the availability and affordability of graphene may help slow the increase in atmospheric carbon dioxide levels via CCUS and accelerate the transition to renewable energy. Increasing the availability and affordability of graphene can catalyze scientific discoveries, innovation, and economic growth. These innovations will also increase sustainability.

**PROCEDURE**

The procedure for capturing carbon dioxide and converting it into graphite involves 3 steps:
1) Capturing atmospheric carbon dioxide with mordenite
2) Transferring carbon dioxide to the graphite synthesizer
3) Converting carbon dioxide into graphite

**Capturing atmospheric carbon dioxide with mordenite**

The atmosphere primarily consists of nitrogen (~80%) and oxygen (~20%). In 2023, carbon dioxide levels were ~419 ppm or ~0.4% (https://www.nytimes.com/interactive/2024/04/20-/upshot/carbondioxidegrowth.html?searchResultPosition=1; accessed on 8/10/2024). Trace amounts of carbon dioxide can be efficiently trapped by mordenite when air is pumped through it at room temperature, ~293 °K, while nitrogen and oxygen pass through (Figure 2a). Zhou et al. (2021) found that mordenite can adsorb 5.68 mmol $CO_2$/gram of mordenite at ~300 °K and 1 bar and that heating mordenite to 373 °K in a vacuum releases carbon dioxide. Mordenite will be used to illustrate the costs of removing carbon dioxide from the atmosphere for this study (Figure 2a).

**Transferring carbon dioxide to the graphite synthesizer**

After mordenite becomes saturated with carbon dioxide, increasing the temperature of the adsorbent by ~100 °K and lowering the pressure to ~6 torr releases the carbon dioxide (Fig. 2b ). Fig. 3a shows how the evolved carbon dioxide moves to the graphite synthesizer.

**Converting carbon dioxide into graphite**

The graphite reactor converts carbon dioxide into graphite via the Bosch reaction:

$CO_2(g) + 2H_2(g) ===> C(graphite) + 2H_2O(g)$ 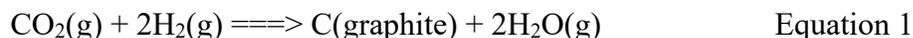 Equation 1

Iron catalyzes this reaction, reaction temperatures range from 450 to 600 °C, and higher pressures favor the forward reaction due to Le Chatelier's principle. Fig 3b. illustrates how hydrogen is added to the graphite synthesizer and Fig. 3c illustrates the production of graphite.

**ECONOMICS OF CONVERTING ATMOSPHERIC CARBON DIOXIDE INTO GRAPHITE**

**Energy cost for carbon dioxide removal and graphite production**

Energy is the greatest expense for the CD2G approach. To estimate the energy costs, I use a value of $0.05/kWh for the cost of electricity, which is based on the $0.049/kWh value for the levelized cost of electricity for new resources entering service in 2027 (United States Energy Information Administration, 2022, Table 1b). This value is for solar power with battery storage and includes transmission.

**Summary of electrical costs for converting carbon dioxide into graphite**

The total electrical costs for converting ~2.5 kg of carbon dioxide into ~0.7 kg of graphite are ~$1.20 (Table 2, Appendices A, B, & C ). These values can be extrapolated to estimate the electrical cost of removing 1 ton of carbon dioxide from the atmosphere:

(~$1.20)/(2.5 kg $CO_2$) = ~$0.48/kg $CO_2$

~$480/ton $CO_2$ at 100% efficiency ((~$0.48/kg) * (1000 kg) = ~$480/ton)

~$640/ton $CO_2$ at 75% efficiency ((~$0.64/kg) * (1000 kg) = ~$640/ton)

~$960/ton $CO_2$ at 50% efficiency ((~$0.96/kg) * (1000 kg) = ~$960/ton)

Using an estimated efficiency for the prototype CD2G factory of ~50% efficiency results in a cost of ~$960/ton carbon dioxide removed from the atmosphere.

**Summary of non-electrical costs for removing carbon dioxide from the atmosphere**

Table 3 summarizes estimates for the non-electrical cost of building a prototype factory to convert carbon dioxide into graphite. 2.5 kg of adsorbent can trap 7300 kg of $CO_2$/year (Appendix D). Using twice as much adsorbent would double the amount of carbon dioxide removed from the atmosphere. The non-electrical costs would be about $12,200 (Table 3). This can be spread out over 30 years, the estimated lifetime of a CD2G plant:

~$12,200/((~7.3 tons)*(~30 years)) = ~$56/ton.

For a prototype factory, the estimated non-electrical costs are ~$56/ton.

**Profits generated by CD2G prototype factory**

The CD2G prototype factory will produce graphite and oxygen. The graphite can be sold for ~$478/ton of $CO_2$ removed. The oxygen can be sold for ~$182/ton of $CO_2$ removed.

*Graphite*

1,000 kg of $CO_2$ can be converted into ~273 kg graphite:

(1000 kg $CO_2$) * (~12 grams C/~44 grams $CO_2$) = ~273 kg C (graphite).

The graphite produced by this method can be sold for ~$500/ton to ~$2,300/ton generating ~$137 to ~$628 for each ton of $CO_2$ captured (Table 4). I use ~$478/ton $CO_2$ removed for a prototype factory.

*Oxygen*

For every ton of carbon dioxide removed from the atmosphere, 736 kg of oxygen will be produced (Appendix E). The oxygen is produced when water is split. Oxygen has a market value of ~$570/ton: (https://ycharts.com/indicators/us_producer_price_index_industrial_gas_manufacturing_oxygen_yearly; accessed on 8/14/24).

Hospitals are converting electricity into hydrogen and oxygen via hydrolysis using the resulting hydrogen for fuel cells and oxygen for patients. These hospitals include the Rijnstate hospital of Elst in the Netherlands and Viamed San José Hospital in Spain (https://undecidedmf.com/why-hydrogen-does-have-a-future/; accessed on 12/28/23).

736 kg of $O_2$ is worth ~$420. Compressing, purifying, and transporting oxygen would decrease the value of the oxygen produced. The cost of compressing 736 kg of oxygen is ~$238 (Appendix F). So the net value of oxygen produced would decrease to ~$182/ton of captured carbon dioxide (Table 4).

*Carbon credits*

Carbon credits may also be a source of profits. These credits were not included in my calculations to make the profit estimates more conservative.

*Summary of Costs using Prototype Technology*

Removing a ton of carbon dioxide from the atmosphere would produce ~$660 worth of graphite and oxygen (Table 4). The electrical and non-electrical costs for removing a ton of carbon dioxide from the atmosphere total ~$1016/ton of carbon dioxide removed from the atmosphere (Table 4). The prototype CD2G factory will remove a ton of carbon dioxide from the atmosphere for a net cost of ~$356/ton (Table 4).

**FROM PROTOTYPE TO OPTIMIZED CD2G FACTORY**

The prototype factory will be used as an experimental platform to increase the efficiency and profitability of the CD2G process. For example, the graphite synthesizer will be optimized to produce the largest (most valuable) flakes. The best configuration, temperatures, flow rates, $H_2/CO_2$ ratios, and pressures will be determined. Also, the ideal composition, texture, shape, and charge of the catalyst will be identified. If the optimized graphite catalyzer produces the largest flakes, one could sell ~$628 of graphite for every ton of carbon dioxide removed from the atmosphere.

**Decreasing costs: efficiency**

At ~50% efficiency, the prototype factory would require ~$960/ton $CO_2$ removed from the atmosphere. As the technology matures, one expects the efficiency to increase due to Wright's law (Wright, 1936; Nagy et al., 2013) and economies of scale (Robinson, 1958; Stigler, 1958; Scherer, 1980; Pratten, 1991; Morroni, 2006). Wright's law or the learning curve effect was first

observed in the aircraft industry: a doubling of aircraft production caused a 20% decrease in the amount of work needed to build a new aircraft. This law has been found to apply to other industries. One expects that as the production of graphite from CD2G factories increases, the costs will decrease due to the knowledge gained from building and operating multiple factories. Economies of scale law says that increasing the size of a factory will increase the efficiency of production. Due to the learning curve effect and economies of scale, one expects that transitioning to mature factories will increase the efficiency from ~50% to ~75%. This would drop the cost of removing a ton of $CO_2$ from the atmosphere to ~$640/ton $CO_2$. If the cost of renewable energy drops (see below), this will further decrease the cost of converting atmospheric carbon dioxide into graphite.

**Decreasing costs: electrical**

Over time, the cost of renewable energy should decrease, due to Wright's law and economies of scale, as more and larger solar power plants, wind farms, and renewable energy storage facilities come online. Swanson's Law, which resembles Wright's law, states that solar panel costs are expected to decrease by 20% every time the cumulative shipped volume doubles (Swanson, 2006). In 2006, solar modules that could produce 320,208 of peak kW were purchased in the US; in 2022, that number grew to 31,679,435. (https://www.eia.gov/renewable/annual/solar_photo/pdf/pv_table3.pdf). Wind power has also grown rapidly since 2000, driven by research and development, supportive policies, and falling costs. Worldwide wind generation capacity has increased, jumping from 7.5 GW in 1997 to 733 GW in 2018 (https://www.irena.org/Energy-Transition/Technology/Wind-energy). The cost of solar energy has decreased from $0.417/kilowatt-hour in 2010 to $0.048/kilowatt-hour in 2021 (Osman et al., 2023). During the same time, the cost of onshore wind decreased by 68%, the cost of offshore wind decreased by 60%, and concentrated solar likewise decreased by 68% (Osman, et al., 2023). As renewable energy prices decrease, demand increases. In turn, producing additional renewable energy capacity decreases the price of renewable energy, resulting in a synergy between decreasing prices and increasing demand.

Solar and wind energy must be stored, which adds expense and decreases the efficiency of solar and wind power. One way to store energy is with lithium batteries. The cost of lithium batteries is projected to decrease in the future and follow a similar trend as the cost of solar energy. For example, Ziegler & Trancik (2021) found that the cost of lithium batteries has dropped by 97 percent since 1991. Pumped hydro energy storage, flow batteries, iron-air batteries, and thermal storage may prove less expensive than lithium ion batteries for long-term energy storage (Kunzig, 2024). Given the projected cost decrease in renewable energy and lithium batteries, a 40% decrease in electrical costs in ~10 years seems reasonable. Decreasing the electrical rate from 0.05/kwh to 0.03/kwh (i.e., a 40% decrease), will decrease CD2G electrical costs from ~$640/ton $CO_2$ captured to ~$384/ton $CO_2$ captured (Tables 4 & 5). Furthermore, as the CD2G technology matures, electrical consumption will decrease. For example, the blower requires 6.5 hp to move 1,121 cubic feet per minute (cfm) of air. A different approach may replace the blower with an industrial fan, which can move 5,000 cfm using 0.3 hp. The efficiency of the other pumps used for the CD2G approach will also improve.

**Decreasing costs: non-electrical**

Non-electrical costs will decrease as the technology matures due to economies of scale (see above) and Wright's Law (see above). If costs drop by 20%, then the non-electrical costs will drop to ~$45/ton $CO_2$ removed:

(~$56/ton $CO_2$) * (0.8) = ~$45/ton $CO_2$ removed (Table 5)

One expects the non-electrical costs to drop by at least 20% for a mature C2G factory.

**Increasing revenue from selling graphite and oxygen**

The graphite synthesizer can be optimized to produce graphite flakes that are larger than 32 mesh, boosting the price of graphite from ~$1750/ton graphite to ~$2300/ton graphite (Table 1). The demand and price for graphite is expected to increase (Ballinger et al., 2019). The increase in graphite demand may shift the price of 32 mesh graphite to greater than ~$3000/ton. The profit generated by selling oxygen will be the same for both the prototype and mature factory.

**Net profit for optimized CD2G factory**

Table 5 summarizes the costs and profits of an optimized CD2G factory.

*Cost summary*:

electrical costs: (~$640/ton $CO_2$) * (0.6) = ~$384/ton $CO_2$ (electricity costs decrease by 40%)

non-electrical costs: (~$56/ton $CO_2$) * (0.8) = ~$45/ton $CO_2$ (economies of scale)

*total cost*: ~$429/ton $CO_2$

*Profit summary*:

graphite     (0.273 tons) * (~$2300/ton) = ~$628
oxygen     ~$182

*gross profit*: ~$810/ton $CO_2$

*Net profit*: (~$810) – (~$429) = ~$381/ton $CO_2$ removed

Removing atmospheric carbon dioxide for a profit provides a strong incentive for decreasing atmospheric carbon dioxide levels. Increasing the availability and affordability of graphite will decrease the cost of energy storage, possibly accelerating the transition to renewable energy.

**SUSTAINABILITY**

The CD2G approach is sustainable at the GtC/yr level. The sustainability of the CD2G approach depends on the amount and type of energy used, water availability, the ecological footprints of the adsorbents, catalysts, and the chemical and mechanical infrastructure.

*Energy*

CD2G factories will be powered by solar and wind energy, which are among the most sustainable energy sources. CD2G factories will increase graphite availability, which will increase the availability and affordability of energy storage, and increase graphene availability, which will catalyze the development renewable energy and sustainability technology. For example, many batteries and fuel cells use use large amounts of graphite. Accelerating the development of renewable energy storage is key for decreasing atmospheric carbon dioxide levels.

*Water availability*

The amount of water needed for CD2G factories is small. For every mole of carbon dioxide removed from the atmosphere, two moles of hydrogen are required. Further, water will be regenerated when graphite is produced (Equation 1). Also, seawater can be used to generate the hydrogen (Seenivasan et al., 2024). Water should not be limiting.

*Adsorbents*

The amount of adsorbent used for the CD2G approach is small. For example, 10 kg of adsorbent can remove up to 7.3 tons of carbon dioxide a year. Adsorbents typically last decades. Further, the ecological footprint of the mordenite adsorbent is small. Zhou et al. (2021) concluded that synthesizing mordenite is economical, uses little energy, and is environmentally benign. Lin et al. (2021) produced a metal-organic framework adsorbent with properties similar to mordenite: it has a small ecological footprint and is durable. New adsorbents with better performance and smaller environmental footprints may be developed as this field matures. In short, the use of carbon dioxide adsorbents is both environmentally and economically sustainable.

*Catalysts for hydrogen production*

The CD2G factory uses catalysts to split water to produce hydrogen and to convert carbon dioxide into graphite. Catalysts for splitting water that have low costs and small ecological footprints are being developed (Chen et al., 2021; Wang et al., 2021; Hodges et al., 2022; Gao et al., 2023). For example, Shiokawa et al. (2024) developed metal anodes for neutral seawater electrolysis that do not contain noble metals. Also, Seenivasan et al. (2024) developed a nickel-tungsten nitride (Ni-W5N4) alloy to split seawater more efficiently without the use of exotic materials. Increasing the efficiency of the electrolyzer decreases the amount of energy needed to split water to produce hydrogen and oxygen and will lower the costs associated with making graphite from atmospheric carbon dioxide.

*Catalysts for graphite production*

Iron can be used as a catalyst to convert carbon dioxide into graphite (Equation 1). A kg of iron can convert 7.3 tons of carbon dioxide into graphite in a year. This catalyst may last for decades. Iron is inexpensive, abundant, and can be readily recycled. It would be hard to imagine a more sustainable catalyst.

*Mechanical infrastructure*

Most of the mechanical infrastructure, such as cement and steel, required for CD2G factories has a small ecological footprint, a long lifetime, and can be recycled at end of use.

**Using Direct Air Capture to remove $CO_2$ from air instead of Flue Gas Capture**

While some Carbon Capture Storage and Utilization (CCSU) approaches use Flue Gas Capture (FGC), these approaches will not be useful after fossil fuel becomes a neglible energy source. In contrast, the CD2G approach described here will remain effective because it uses Direct Air Capture (DAC) to remove carbon dioxide from the atmosphere. This approach will work when fossil fuel is no longer burned.

**Scalability**

In addition to being sustainable, CD2G technologies can be scaled up in size and quantity to remove gigatons of carbon dioxide every year.

*Energy*

Energy is the largest resource needed to convert atmospheric carbon dioxide into graphite. Using the CD2G approach to remove 1 Gt. $CO_2$ would require ~$1.3 \times 10^{13}$ kWh energy using a 75% efficiency (see table 2):

(~32 kWh/2,500 g $CO_2$) * ($1 \times 10^{15}$ grams $CO_2$/Gt $CO_2$) = ~$1.3 \times 10^{13}$ kWh.

In 2009, Lu et al. estimated that "land-based 2.5-megawatt (MW) turbines restricted to nonforested, ice-free, nonurban areas operating at as little as 20% of their rated capacity could supply >40 times current worldwide consumption of electricity, >5 times total global use of energy in all forms" (10933). Global onshore and offshore wind power potential is $8.7 \times 10^{17}$ kWh/yr (https://www.nrel.gov/docs/fy17osti/65323.pdf). De Castro et al. (2013) estimated the global capacity for solar energy to range from 1.7 to $3.3 \times 10^{13}$ kWh/yr. Dupont et al. (2020) estimated the global capacity for solar energy to range from $5.9 \times 10^{14}$ to $3.9 \times 10^{15}$ kWh/yr. There is sufficient renewable energy potential to power the CD2G approach at the Gt $CO_2$/yr scale.

*Non-energy resources*

Most materials needed to build Gt/yr scale CD2G plants are not limiting, such as cement, steel, pipes, valves, and computers. These materials are inexpensive, readily available, and represent proven and robust technology. To increase the efficiency of the pumps, high efficiency motors may be used. These would decrease the amount of energy required to convert atmospheric carbon dioxide into graphite. High efficiency motors are being developed that don't use rare materials (Riba et al., 2016).

Only a small amount of catalysts are needed for CD2g factories. These include iron and mordenite. Iron is abundant and synthesizing mordenite is economical, energy efficient, and environmentally benign (Zhou et al., 2021).

**Siting**

CC2G factories will be most effective if sited near sources of inexpensive and abundant renewable energy and where there is a demand for graphite and oxygen.

**Synergies**

Increasing the availability and affordability of graphite will help accelerate the transition to renewable energy because graphite is needed for some thermal batteries, lithium ion batteries and fuel cells. For example, some thermal batteries use carbon blocks (Ramkumar & Patterson, 2024). Carbon blocks can be made from graphite. Many electric-vehicle battery electrodes and some large-capacity battery electrodes use graphite (Gastol et al., 2021). Switching to battery electric vehicles that are charged by renewable energy and using large-capacity batteries to store renewable energy on a grid scale will decrease the amount of carbon dioxide released to the atmosphere by burning fossil fuel. The increased availability and affordability of graphite will lower the cost of renewable energy, which, in turn, will lower the cost of removing carbon dioxide from the atmosphere and converting it into graphite using the CD2G method.

Graphite electrodes are also used for fuel cells (Rao et al., 2020). For example, some fuel cells use graphite for their bipolar plates, gas diffusion layers, or catalysts (https://www.innovation-newsnetwork.com/graphite-hydrogen-fuel-cell-technologies/26736/; 8/24/24). Fuel cells may propel cars, trucks, trains, planes, and store energy at scales ranging from residential to grid. Storing renewable energy in fuel cells can lead to decreased carbon dioxide emissions from fossil fuel combustion.

In a sustainable feedback loop, increasing the availability and affordability of graphene will increase the effectiveness of catalysts used for generating hydrogen from water and the catalysts used for converting hydrogen into energy. More effective catalysts will decrease the cost of renewable energy. The lower energy cost will lower the cost of converting atmospheric carbon dioxide into graphite, which will, in turn, increase the affordability and availability of graphene. Graphene-copper wires exhibit a 450% increase in electrical current carrying capacity, a 41% higher electrical current and a 224% thermal heat dissipation increase compared to pure copper (https://mae.osu.edu/events/2023/03/can-graphene-based-electrical-conductors-replace-copper;

8/24/24). Most of the wires used for electrical grids are made out of aluminum, which has a lower conductivy compared to copper. Replacing existing power lines in the grid with graphene-copper wires would greatly increase the capacity of the grid to absorb and distribute renewable energy.

**Discussion**

Burning fossil fuel has been linked with climate change, air pollution, water pollution, solid waste production, and health problems (Miller & Spoolman, 2009). Increased atmospheric carbon dioxide levels causes ocean acidification (Harrison, 2019; https://oceanservice.noaa.gov-/facts/acidification.html; 02/02/25). The transition to new forms of energy is driven by economics and convenience. The transition from fossil fuel energy to renewable energy will be rapid when the price of renewable energy is less than half the price of fossil fuel energy and when renewable energy is easier to use. The primary goal of this research is to decrease the cost of renewable energy by decreasing the cost of graphite. The secondary goal of this research is to remove carbon dioxide from the atmosphere for a profit. The mature CD2G factory accomplishes both of these goals. Finally, CD2G factories may stimulate innovation and the economy by making graphite and graphene more affordable and available.

The technology for creating CD2G factories is mature and robust. A prototype factory could be built in a year. This factory could be used to optimize the CD2G process and the results could lead to mature CD2G factories in 2 to 3 years. A CD2G factory is analogous to a gas chromatograph.

China, India and Brazil mine most of the world's graphite: from 2015 to 2016, China produced 66%, India produced 14%, and Brazil produced 7% of the world's graphite (Jara et al., 2019). Graphite is needed for industry and is important for national security (Jara et al., 2019). CD2G factories can be built in any country, possibly reducing economic and political tensions associated with a high-demand substance that is produced mainly by a few countries.

Mining typically often produces solid mine waste and acid mine waste, releases radioactive materials to the environment, causes landscape degradation, and creates air and water pollution (Miller & Spoolman, 2009). If CD2G factories produce graphite for a lower cost than graphite mining, they could eliminate graphite mines and the ecological, economic, and political problems associated with graphite mining.

It is possible that the supply of graphite will exceed the demand if CD2G factories operate at the gigaton level for an extended time. At this point, storing the excess graphite at a cost of $429/ton (Table 5) may be preferable to paying the cost of $100,000/ton (Archer et al., 2020) for leaving the carbon dioxide in the atmosphere.

**Conclusion**

Increasing levels of atmospheric carbon dioxide may cause catastrophic damage. This cost may be as high as $100,000/ton of carbon dioxide released to the atmosphere (Archer et al., 2020).

The Carbon Dioxide to Graphite (CD2G) approach converts atmospheric carbon dioxide to graphite, which will slow the build-up of atmospheric carbon dioxide levels, which will slow climate change and ocean acidification. A mature CD2G factory could remove atmospheric carbon dioxide for a net profit of ~$381/ton. This approach will increase the availability and affordability of graphite. Since graphite is essential for many thermal and lithium batteries and fuel cells, increasing the supply and affordability of graphite will decrease the cost of storing renewable energy. In turn, this will accelerate the transition to renewable energy, decreasing the release of carbon dioxide to the atmosphere by fossil fuel combustion. Graphite can be converted to graphene, which will increase the affordability and availability of graphene, and offer opportunities for additional synergies, scientific advances, technological progress, and economic opportunities, including increased efficiencies of energy generation, energy storage, and energy use. CD2G factories are sustainable and can be scaled to remove atmospheric carbon dioxide at the Gt level.

**Acknowledgements**

I thank BethAnn Zambella. This paper would not be possible without her encouragement and support. She also helped with revising, editing, proofreading, and taming the references. I thank Eric Smith and Lori Weeden for reviewing the manuscript.

Fig. 1. Converting atmospheric $CO_2$ into graphite: schematic overview.

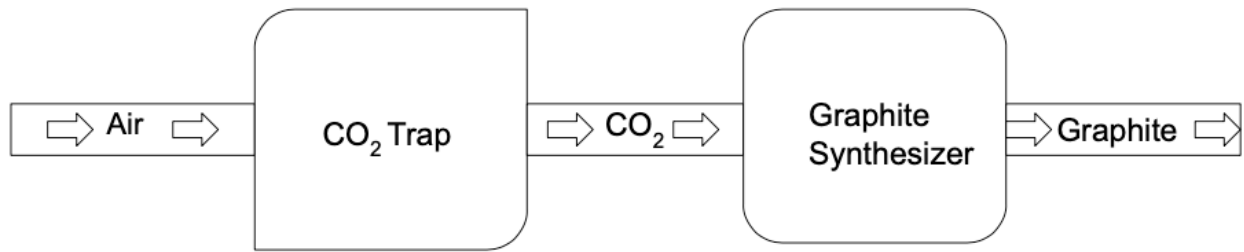

$$CO_2 + 2H_2 ===> graphite [C] + 2H_2O$$

Atmospheric carbon dioxide is removed from the atmosphere with a $CO_2$ trap. After being released from the trap by heating, the carbon dioxide is converted into graphite using the Bosch reaction.

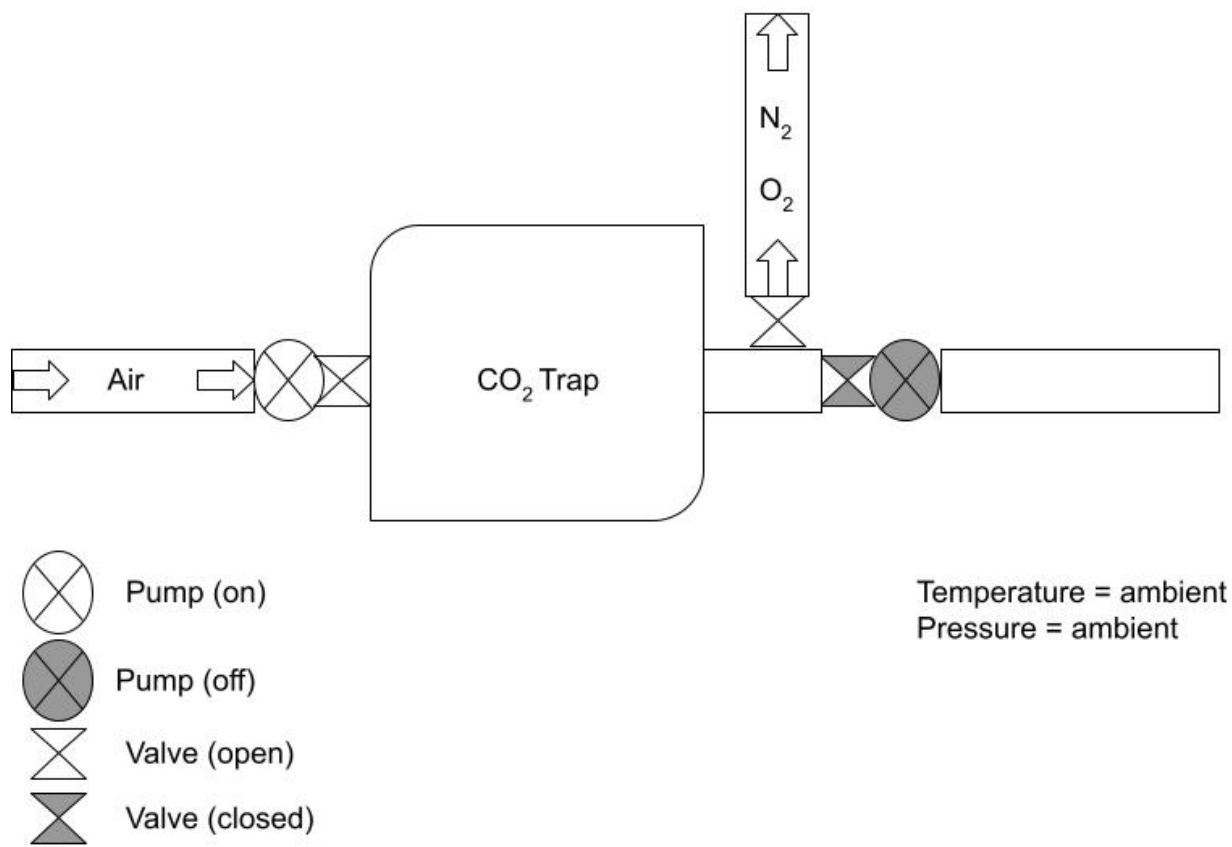

Fig. 2a. $CO_2$ capture.

The atmosphere is pumped through a $CO_2$ trap filled with an adsorbent. Carbon dioxide is trapped, while nitrogen and oxygen are released to the atmosphere.

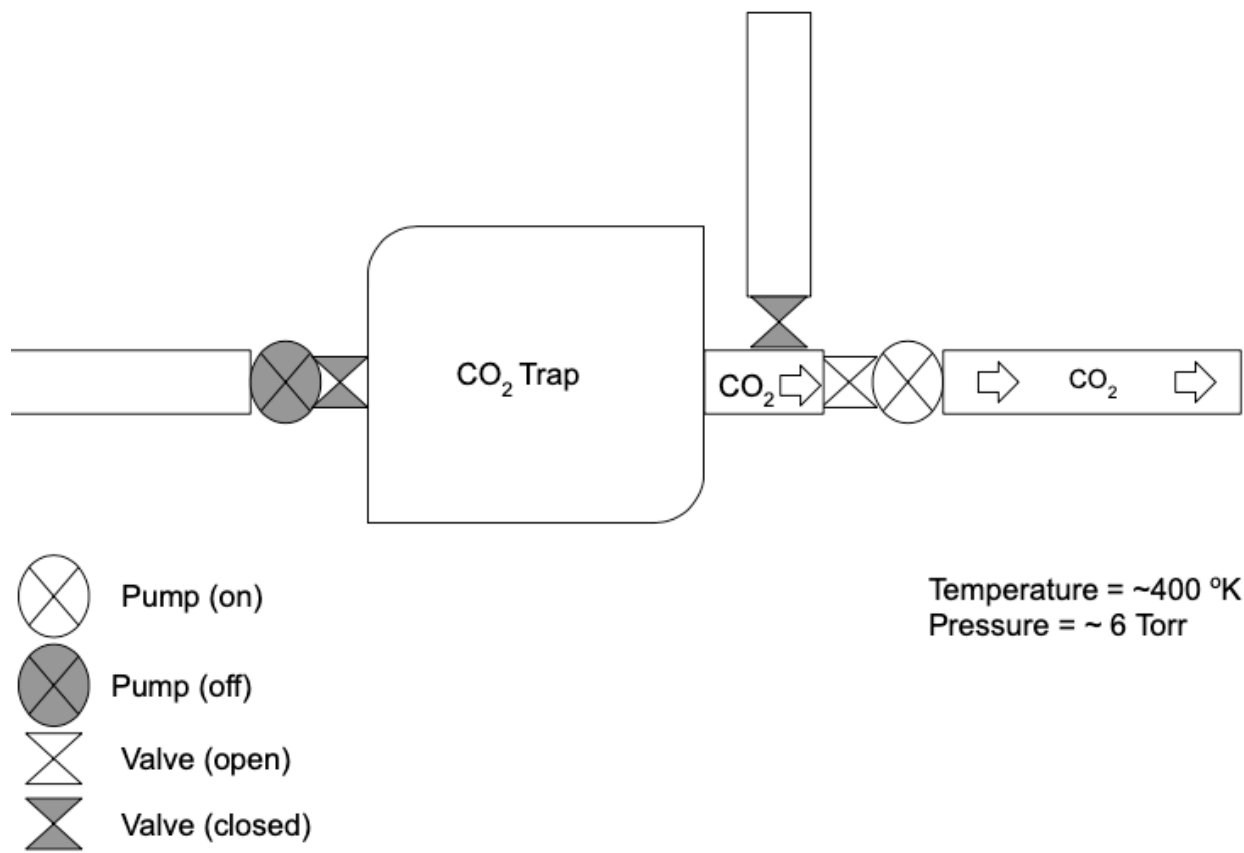

Fig. 2b. CO$_2$ transfer.

Carbon dioxide is released from the trap by heating it to ~400 °K and in a vacuum of ~6 Torr. The evolved carbon dioxide is transferred to the graphite reactor.

Fig. 3a. Graphite synthesis: loading $CO_2$ into graphite synthesizer.

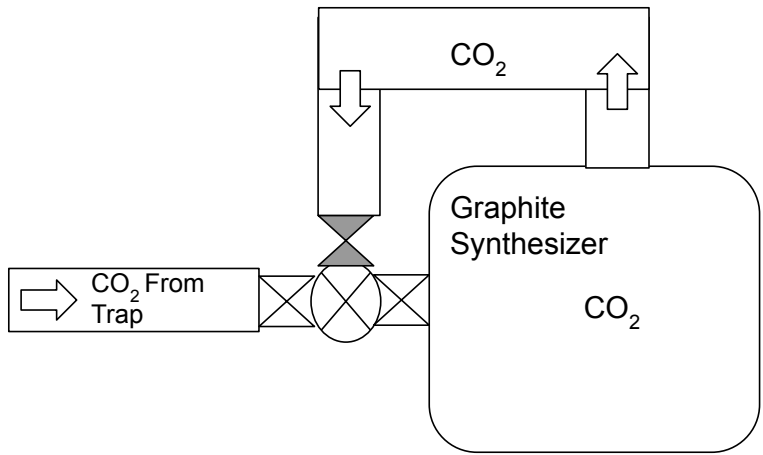

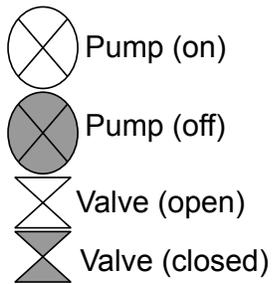

Temperature: ambient
Pressure: elevated

Carbon dioxide is loaded into the graphite synthesizer.

Fig. 3b. Graphite synthesis: loading H$_2$ into graphite synthesizer

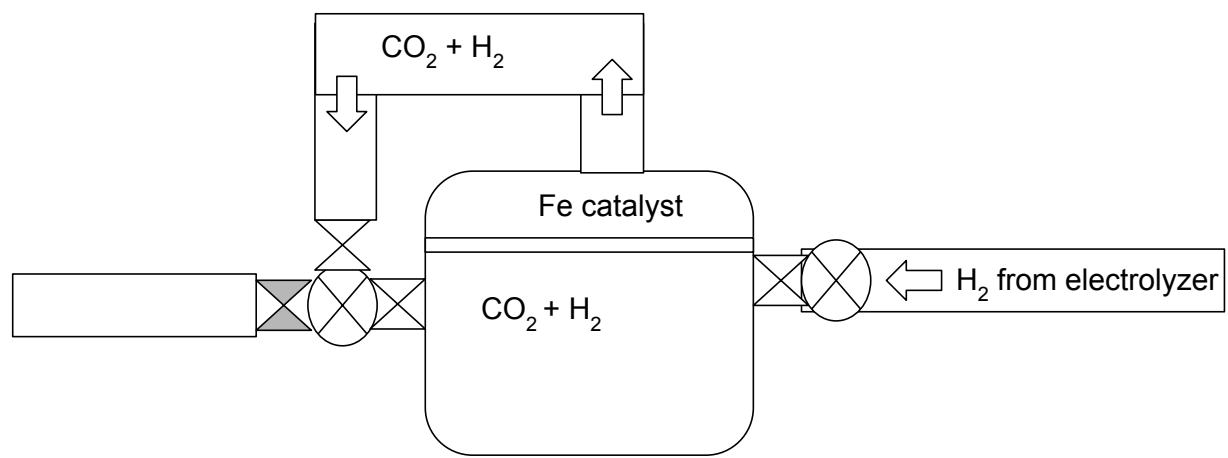

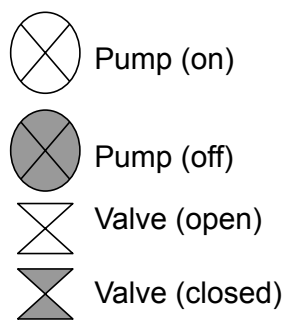

Pump (on)

Pump (off)

Valve (open)

Valve (closed)

Temperature: ambient
Pressure: elevated

Hydrogen is added to the carbon dioxide in the graphite reactor. The hydrogen is produced by an electrolyzer.

Fig. 3c. Graphite synthesis: production

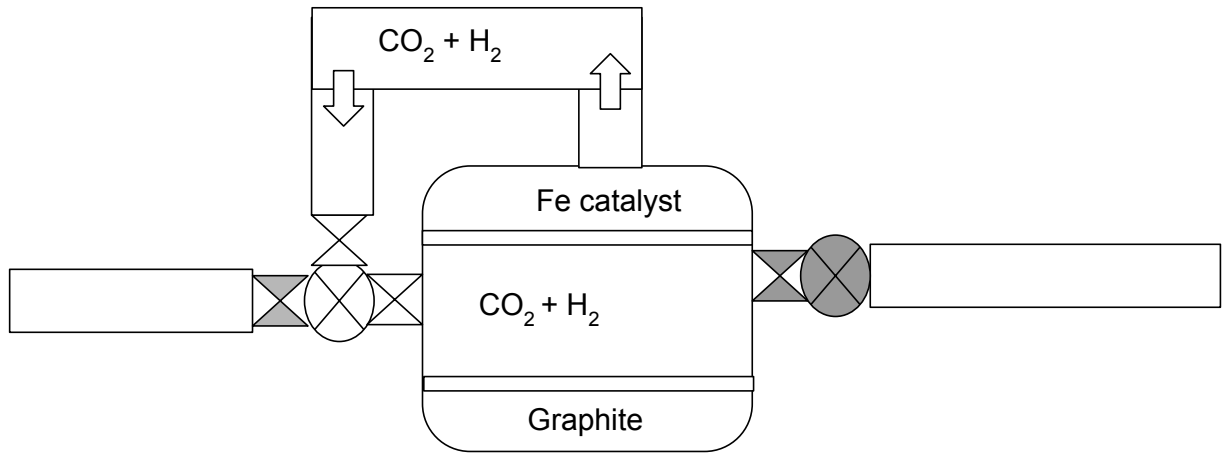

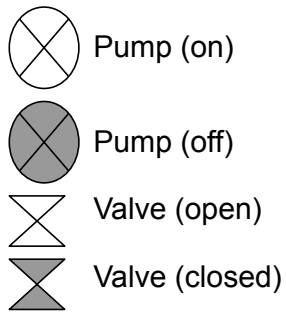

Pump (on)

Pump (off)

Valve (open)

Valve (closed)

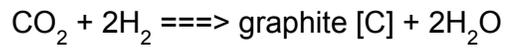

$CO_2 + 2H_2 ===>$ graphite [C] $+ 2H_2O$

Temperature: high
Pressure: high

After the carbon dioxide and hydrogen have been added to the graphite reactor at high pressure, the temperature is increased. The iron in the graphite reactor catalyzes the $CO_2$-to-graphite conversion. High pressure favors the formation of graphite.

Table 1. Graphite prices; larger pieces of graphite command higher prices (Northern Graphite, https://www.northerngraphite.com/about-graphite/graphite-pricing/, accessed 10/14/24)

| Graphite size | Price (graphite produced) | Price ($CO_2$ removed) |
| --- | --- | --- |
| +32 mesh | $2,300/ton | $628/ton $CO_2$ |
| +50 mesh | $1,750/ton | $478/ton $CO_2$ |
| +80 mesh | $1,400/ton | $382/ton $CO_2$ |
| +100 mesh | $1,200/ton | $328/ton $CO_2$ |
| +150 mesh | $800/ton | $218/ton $CO_2$ |
| -150 mesh | $500/ton | $137/ton $CO_2$ |

The price per ton of $CO_2$ removed is calculated by multiplying the price per ton of graphite by 0.273 (12 grams C/44 grams $CO_2$). An optimized CD2G factory will make mostly large flakes (+32 mesh size). To achieve this, the optimum configuration, temperatures, flow rates, $H_2/CO_2$ ratios, and pressures will be determined. Also, the optimum composition, texture, shape, and charge of the catalyst will be determined. See costs and expenses in Tables 2-5.

Table 2. Approximate electrical costs for converting 2.5 kg of $CO_2$ to 0.7 kg of graphite

| Task | Cost 100% efficient | 75% efficient | 50% efficient |
|---|---|---|---|
| Pump air through Adsorbent (1) | $0.40 | $0.53 | $0.80 |
| Heat adsorbent to release $CO_2$ (2) | $0.01 | $0.01 | $0.02 |
| Transfer $CO_2$ to Graphite Synthesizer (2) | $0.10 | $0.13 | $0.20 |
| $H_2$ generation (3) | $0.48 | $0.64 | $0.96 |
| Pump $H_2$ to Graphite Synthesizer (3) | $0.10 | $0.13 | $0.20 |
| Heat Graphite Synthesizer (3) | $0.01 | $0.01 | $0.02 |
| Pump $CO_2$ and $H_2$ mix through Graphite Synthesizer (3) | $0.10 | $0.13 | $0.20 |
| Total | ~$1.20 | ~$1.60 | ~$2.40 |

Pumping air through the adsorbent and splitting water to generate hydrogen use the bulk of the electricity. Finding a more efficient way to get the adsorbent saturated with carbon dioxide is one way to reduce costs. (1) Appendix A. (2) Appendix B. (3) Appendix C.

At 100% efficiency, it will take 24 kWh of electricity to trap 2.5 kg of $CO_2$:
($1.20/2.5 kg $CO_2$) * (1 kWh/$0.05) = 24 kWh/2.5 kg

At 75% efficiency, it will take 32 kWh of electricity to trap 2.5 kg of $CO_2$:
($1.60/2.5 kg $CO_2$) * (1 kWh/$0.05) = 32 kWh/2.5 kg

At 75% efficiency, it will take 48 kWh of electricity to trap 2.5 kg of $CO_2$:
($2.40/2.5 kg $CO_2$) * (1 kWh/$0.05) = 48 kWh/2.5 kg

Table 3. Carbon trap and graphite synthesizer non-electrical costs

| Item | System Cost |
|---|---|
| Positive displacement blower | ~$2,000 |
| Vacuum pump | ~$2,600 |
| Circulator pump | ~$2,600 |
| Pressure gauges, heaters, tubing, valves | ~$1,000 |
| $CO_2$ analyzer | ~$2,000 |
| $H_2$ generator | ~$2,000 |
| Total | ~$12,200 |

Table 4. Prototype expenses/profits (all values are approximations)

|  | range | ~$/ton $CO_2$ captured |
|---|---|---|
| **Expenses**: | | |
| electrical | 480-960 | 960 |
| non-electrical |  | 56 |
| **total** |  | **1016** |
| **Profits**: | | |
| oxygen |  | 182 |
| graphite | 137-628 | 478 |
| **total** |  | **660** |
| **Net cost** |  | **356** |

Based on Tables 1-3, the prototype CD2G experiment can remove atmospheric carbon dioxide at a cost of ~$356/ton. See Table 5 for profits as the technology matures.

Table 5. Expenses/profits for mature technology (all values are approximations)

|  | $/ton $CO_2$ captured |
|---|---|
| **Expenses**: | |
| electrical | ~384 |
| non-electrical | ~45 |
| **total** | **~429** |
| **Profits**: | |
| graphite | ~628 |
| oxygen | ~182 |
| **total** | **~810** |
| **Net profit** | **~381** |

A mature CD2G factory will generate a ~$381/ton profit for removing a ton of atmospheric carbon dioxide.

technology improvement and cost decline. *Energy Environ. Sci.*, 14, 1635.

**Appendix A. Cost of trapping 2.5 kg of atmospheric carbon dioxide**

All of the values below are approximations. The mordenite adsorbent can remove 5.68 mm $CO_2$/g of adsorbent (Zhou et al., 2021). So, 10,000 g (10 kg) of adsorbent can remove 56.8 moles of $CO_2$:

(10,000 g adsorbent) * (5.68 x $10^{-3}$ moles $CO_2$/g adsorbent) = 56.8 moles of $CO_2$.

Amount of air needed to capture 56.8 moles moles of $CO_2$ is about 1.14 x $10^5$ cubic feet:

(1 mole air/0.0004 moles $CO_2$) * 56.8 moles $CO_2$ = 1.42 x $10^5$ moles air

The contemporary atmospheric carbon dioxide value is ~400 ppm (i.e., 0.000400 moles $CO_2$/mole air).

Convert to liters:

22.4 liters ~ one mole of air

1.42 x $10^5$ moles air * (22.4 liters of air/mole of air) = 3.18 x $10^6$ liters

Convert to cubic feet:

28 liters ~ 1 cubic foot

3.18 x $10^6$ liters * (1 cubic foot/28 liters) = 1.14 x $10^5$ cubic feet

One way to move air through the adsorbent is with a 6.5 hp blower that moves ~1121 cfm. It will take about 100 minutes to blow enough air (1.14 x $10^5$ cubic feet) to capture 56.8 moles of carbon dioxide:

1.14 x $10^5$ cubic feet /1121 cfm = 100 minutes

100 minutes = 1.7 hours

It would cost about $0.40 to capture 56.8 moles of $CO_2$:

Convert horsepower to kW:

1 hp ~ 0.75 kW

(6.5 hp) * (0.75 kW/hp) = 4.9 kW

4.9 kW x 1.7 hours = 8.3 kWh

8.3 kWh x $0.05/kWh = $0.40 to adsorb 56.8 moles of $CO_2$

(56.8 moles $CO_2$) * (44 grams $CO_2$/mole) = 2.5 kg

$0.40/2.5 kg = $0.16/kg $CO_2$

($0.16/kg $CO_2$) x (1000 kg/ton) = $160/ton $CO_2$ removed

It will cost ~$160 to capture a ton of carbon dioxide.

How much electricity is needed to capture a ton of carbon dioxide:

(8.3 kWh/2.5 kg) * 1000 kg = 3000 kWh

The largest cost associated with the CD2G approach is energy. Energy costs will be about $0.05/kWh. This is based on an estimate of $0.049/kWh for solar hybrid energy (US EIA, 2022, Table 1b). This cost may decrease to $0.03/kWh over time as renewable energy and energy storage technology mature, greatly cutting costs.

**Appendix B. Cost of desorbing carbon dioxide and moving it to graphite reaction chamber.**
This back-of-the-envelope estimate is derived from approximations.

*It will cost about $0.01 to heat 10 kg of adsorbent by ~100 °K:*

$\Delta Q = mc\Delta T$

$\Delta Q$ = change in heat

m = 10 kg (mass)

c = 830 J/(kg °K) specific heat (using quartz sand as best guess)

$\Delta T$ = 100 °K

$\Delta Q$ = (10 kg) * (830 J/(kg °K) * (100 °K) = 8 x $10^5$ J

Convert J to kwh:

1 kWh = 3.6 x $10^6$ J

8.0 x $10^5$ J * (1 kWh/3.6 x $10^6$ J) = 0.2 kWh to release 2.5 kg of $CO_2$ from 10 kg of adsorbent

$0.05/kWh

0.2 kWh * $0.05/kWh = $0.01 to release 2.5 kg of $CO_2$

Power needed to release a ton of $CO_2$:

(0.2 kWh/2.5 kg CO2) * (1000 kg/ton) = 80 kWh

Cost needed to release a ton of $CO_2$:

80 kWh * 0.05 = $4.00

*The cost of pumping 56.8 moles of carbon dioxide into the Bosch reaction chamber will be about $0.10*:

Volume of $CO_2$:

(56.8 moles $CO_2$) * (22.4 liters/mole) = 1.3 x $10^3$ liters

A Fisherbrand™ MaximaDry™ Diaphragm Vacuum Pump (13-880-18) costs ~$2601, can generate a ~6 torr vacuum and can move ~20 liters/minute. The estimated energy consumption is about ~2 kW.

(1.3 x $10^3$ liters)/(20 liter/minute) = 65 minutes = 1.08 hours

1.08 hours * 2kWh = 2.1 kWh

2.1 kWh * 0.05/kWh = $0.10.

**Appendix C. Cost of converting carbon dioxide to graphite. These are approximations.**

*Energy needed to generate enough hydrogen to convert 57 moles of carbon dioxide into graphite*:

10 kg of adsorbant can capture ~57 moles of carbon dioxide (Appendix A).

Two moles of hydrogen are needed to convert one mole of carbon dioxide into one mole of graphite ($CO_2 + 2H_2 ===> C(graphite) + 2H_2O$); (Equation 1).

Converting ~57 moles of $CO_2$ to graphite requires ~114 moles $H_2$.

Splitting a mole of liquid water to produce a mole of hydrogen at 25°C requires 285.8 kJ of energy—237.2 kJ as electricity and 48.6 kJ as heat.

*It will take 34,200 kJ of energy to produce 114 moles of $H_2$*:

(114 moles $H_2$) * (~300 kJ/mole $H_2$) = 34,200 kJ

*Since one kWh equals 3600 kJ, it will take 9.5 kWh at a cost of $0.48 to make 114 moles of hydrogen*:

(34,200 kJ) * (1kWh/3600kJ) = 9.5 kWh

*9.5 kWh costs $0.48*:

(9.5 kWh) * ($0.05/kWh) = $0.48

*Cost of pumping the $CO_2$ and $H_2$ mixture through graphite synthesizer*:

About the same as transferring $CO_2$ into graphite synthesizer, $0.10 (Appendix B)

*Heating the Bosch reactor will cost about $0.01*:

$\Delta Q = mc\Delta T$ (Appendix B)

The iron catalyst for the graphite reactor will weigh 1 kg. The reactor temperature will be increased by ~750 °K and the specific heat (c) of iron is ~450 J/(kg °K).

$\Delta Q$ = (1 kg) * (450 J/(kg °K)) * (750 °K) = 3.4 x $10^5$ J

*Convert J to kWh*:

1 kwh = 3.6 x $10^6$ J

(3.4 x $10^5$ J) * (1 kWh/3.6 x $10^6$ J) = 0.09 kWh

(0.09 kWh) * ($0.05/kWh) = $0.01.

Pumping carbon dioxide and hydrogen through the reactor is about the same cost as pumping $CO_2$ into the reactor: $0.10.

**Appendix D: Estimated amount of carbon dioxide removed in a year using 10 kg of adsorbant**

*It will take about ~100 minutes to trap 2.5 kg of $CO_2$ (Appendix A).*

*It will take about ~65 minutes to release the trapped $CO_2$:*

It will take ~10 minutes to heat 10 kg of adsorbant by 100°C (estimate).

It will take ~55 minutes to desorb the $CO_2$ from the mordenite using a pump (estimate).

*It will take 17 minutes to cool the adsorbant trap with a ~1121 cfm flow rate (estimate).*

*The graphite can be synthesized while the $CO_2$ trap is collecting (estimated at 30 minutes).*

*The total time to process 2.5 kg of $CO_2$ is ~3 hours*:

*~20 kg of carbon dioxide can be removed from the atmosphere in a day or ~7,300 kg in a year.*

**Appendix E. Amount of oxygen produced for every ton of carbon dioxide captured**

*There are 2.3 x 10$^4$ moles of carbon dioxide in a ton of carbon dioxide*:

(1000 kg CO$_2$) * (1000 g/kg) * (1 mole/44 grams) = ~2.3 x 10$^4$ moles carbon dioxide.

It takes 2 moles of hydrogen to convert a mole of carbon dioxide into graphite:

CO$_2$(g) + 2H$_2$(g) ===> C(graphite) + 2H$_2$O(g)     Equation 1

It takes ~4.6 x 10$^4$ moles of hydrogen to convert ~2.3 x 10$^4$ moles of carbon dioxide into graphite.

Splitting water to produce hydrogen will produce one mole of oxygen for every two moles of hydrogen produced:

2H$_2$O ===> 2H$_2$ + O$_2$

*2.3 x 10$^4$ moles (~736 kg) of oxygen will be produced for every ton of carbon dioxide captured*:

(2.3 x 10$^4$ moles oxygen) * (32 grams/mole) * (1kg/1000 grams) = 736 kg O$_2$

**Appendix F. Estimating the cost of compressing oxygen**

*If it costs ~$1 to compress 80 cu. ft. of oxygen, then it will cost ~$238 to compress 736 kg of oxygen*:

Convert 736 kg of oxygen (from Appendix E) to liters:

(736 x $10^3$ grams oxygen) * (1 mole/32 grams) * (22.4 liters/mole) = 5.2 x $10^5$ liters

Convert liters to cubic feet:

5.2 x $10^5$ liters * 1 cu. ft./28 liters = 1.9 x $10^4$ cu. ft.

1.9 x $10^4$ cu ft * $1/80 cu. ft. = $238